\documentclass[notitlepage,11pt,preprint]{article}
\usepackage{authblk}
\usepackage{graphicx}
\usepackage[bitstream-charter]{mathdesign} 
\usepackage[colorlinks=true,urlcolor=blue,linkcolor=blue, citecolor=blue]{hyperref}

\begin{document}

\title{Fermi Problem: Power developed at the eruption of the Puyehue-Cord\'{o}n Caulle volcanic system in June 2011}
\date{}

\author[1,2,3]{H. Asorey\thanks{asoreyh@cab.cnea.gov.ar}}
\author[1,2]{A. L\'{o}pez D\'{a}valos}
\affil[1]{Sede Andina, Universidad Nacional de R\'{\i}o Negro, San Carlos de Bariloche, Argentina.}
\affil[2]{Centro At\'omico Bariloche and Instituto Balseiro, Comisi\'on Nacional de Energ\'ia At\'omica, San Carlos de Bariloche, Argentina.}
\affil[3]{Escuela de F\'isica, Universidad Industrial de Santander, Bucaramanga, Colombia}
\maketitle

\begin{abstract}
  On June 4 2011 the Puyehue-Cord\'{o}n Caulle volcanic system produced a
  pyroclastic subplinian eruption reaching level 3 in the volcanic explosivity
  index. The first stage of the eruption released sand and ashes that affected
  small towns and cities in the surrounding areas, including San Carlos de
  Bariloche, in Argentina, one of the largest cities in the North Patagonian
  andean region.  By treating the eruption as a Fermi problem, we estimated the
  volume and mass of sand ejected as well as the energy and power released
  during the eruptive phase. We then put the results in context by comparing
  the obtained values with everyday quantities, like the load of a cargo truck
  or the electric power produced in Argentina. These calculations have been
  done as a pedagogic exercise, and after evaluation of the hypothesis was done
  in the classroom, the calculations have been performed by the students. These
  are students of the first physics course at the Physics and Chemistry Teacher
  Programs of the Universidad Nacional de R\'{\i}o Negro.\\

  El 4 de Junio de 2011, el sistema volc\'{a}nico Puyehue-Cord\'{o}n Caulle
  produjo una erupci\'{o}n de tipo pirocl\'{a}stica subplinianana que alcanz\'{o} el
  nivel 3 en \'{\i}ndice de explosividad volc\'{a}nica. Durante la primera fase de la
  erupci\'{o}n el sistema produjo emisiones de arenas y cenizas que afectaron a
  pueblos y ciudades cercanas, incluyendo a San Carlos de Bariloche, una de
  las mayores ciudades de la regi\'{o}n Andina, al norte de la Patagonia
  Argentina. Al considerar a la erupci\'{o}n como un Problema de Fermi, hemos
  logrado estimar el volumen y la masa de arena eyectada, as\'{\i} como la energ\'{\i}a
  liberada y la potencia erogada durante la fase eruptiva. Luego, pusimos
  estas cantidades en contexto al compararlas con cantidades presentes en la
  vida diaria, como la carga de camiones o la potencia de la matriz
  energ\'{e}tica de Argentina. Estos c\'{a}lculos fueron realizados como un ejercicio
  pedag\'{o}gico por nuestros estudiantes del primer curso de F\'{\i}sica del
  Profesorado Superior de F\'{\i}sica y de Qu\'{\i}mica de la Universidad Nacional de
  R\'{\i}o Negro. 

\end{abstract}

\section{Introduction}
The renowned Italian physicist Enrico Fermi\,\cite{Fermi,Segre} was famous for
his ability to make reliable estimates, even with data that would have seemed
insufficient to many people. One of the best known examples of this is the
estimate he made of the power of the first atomic bomb detonated on July 16,
1945 at the New Mexico desert, measuring the distance travelled by a few scraps
of paper that he dropped to the ground while participating as an observer of
the explosion\,\cite{Jungk}.

Later, when he served as professor at the University of Chicago, Fermi used to
pose such problems to his students as a teaching method. The most famous was to
estimate the number of piano tuners in Chicago at that time. Making a series of
reasonable assumptions, such as the number of people living in Chicago, the
number of people living on average in every household, every how many houses
there is a piano, how often a piano should be tuned, how long it takes for a
tuner to do his work, etc., he could effectively estimate the number of piano
tuners in Chicago, obtaining a result that compared reasonably well with those
contained in the telephone directory.

In the first year of the Physics Teacher Program at the UNRN we frequently face
the students with everyday problems, which can sometimes be solved ``\`{a} la
Fermi". We find it is a very good method to help students sharpen their
imagination as well as their observation skills, and at the same time is a
training in logical reasoning.

\begin{figure}
\includegraphics[width=0.85\columnwidth]{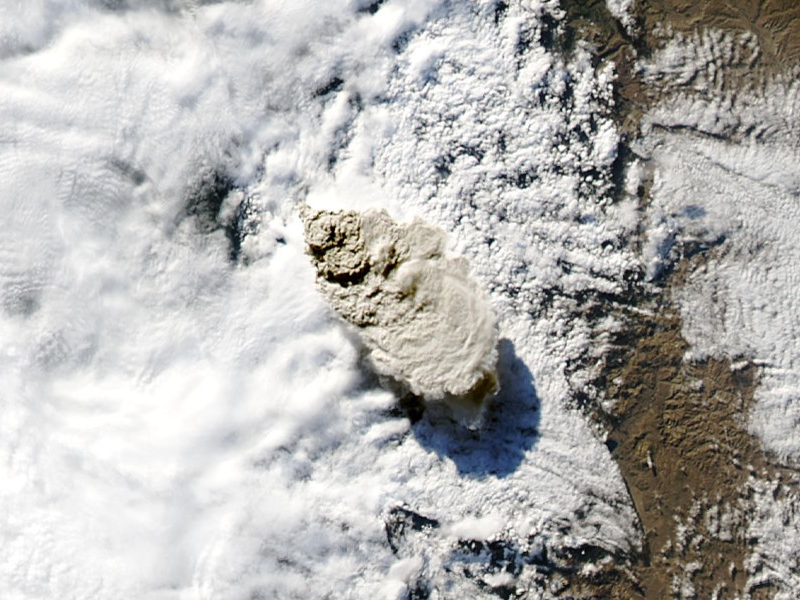}
\caption{Image from NASA Aqua satellite showing the first stages of the
eruption in the Puyehue-Cord\'{o}n Caulle volcanic system, recorded on 04 Jun
2011 at 17:10 UTC, a few minutes after the eruption began.\label{FIaqua}}
\end{figure}

Within this type of problems, we raised the possibility of estimating
the power developed by the Puyehue-Cord\'{o}n Caulle volcanic system, near the
Argentine-Chilean border, in its eruption of June 4, 2011 at 17:00 UTC (14:00
local time, GMT-3). To do this we consider only the first eruption, which
covered with ashes and sand the some small villages in Chile and the cities of
San Carlos de Bariloche ($71.3^\mathrm{o}$\,W, $41.9^\mathrm{o}$\,S), Villa La
Angostura ($71.6^\mathrm{o}$\,W, $40.7^\mathrm{o}$\,S) and surrounding areas in
Argentina. In Figure\,\ref{FIaqua} the first stage of the eruption is shown, as
it has been recorded from space by NASA Aqua satellite\cite{aqua}.

The driving force for this approach was essentially to show the students,
who will be teachers and professors by the end of their studies, that as
educators we must take advantage of every chance we have to teach.

\section{Working Hypothesis}

As in every Fermi problem, the first step is to build a set of reasonable
working hypotheses, starting from personal insights, educated guesses and
general knowledge. Here we resume the our hypotheses as we had found them
during two consecutive lectures.

\begin{enumerate}

\item Area covered by sand and dust.

According to photographs taken by NASA Terra satellite\cite{terra} on June 5,
2011, wind conditions during the eruption day made the cloud of sand and ash
follow almost a straight line running along the Nahuel Huapi lake, in the
WNW-SSE direction. Although the fallout was only partially visible on water, it
is clear that it covered the entire area $A_{NH}$ of the lake. Therefore we use
this area as a parameter to estimate the amount of ash fall. As a working
hypothesis we assume the area covered by ash and sand to be about three times
the area of Nahuel Huapi lake, $A=3 A_{NH}$.

\begin{figure}
\includegraphics[width=0.9\columnwidth]{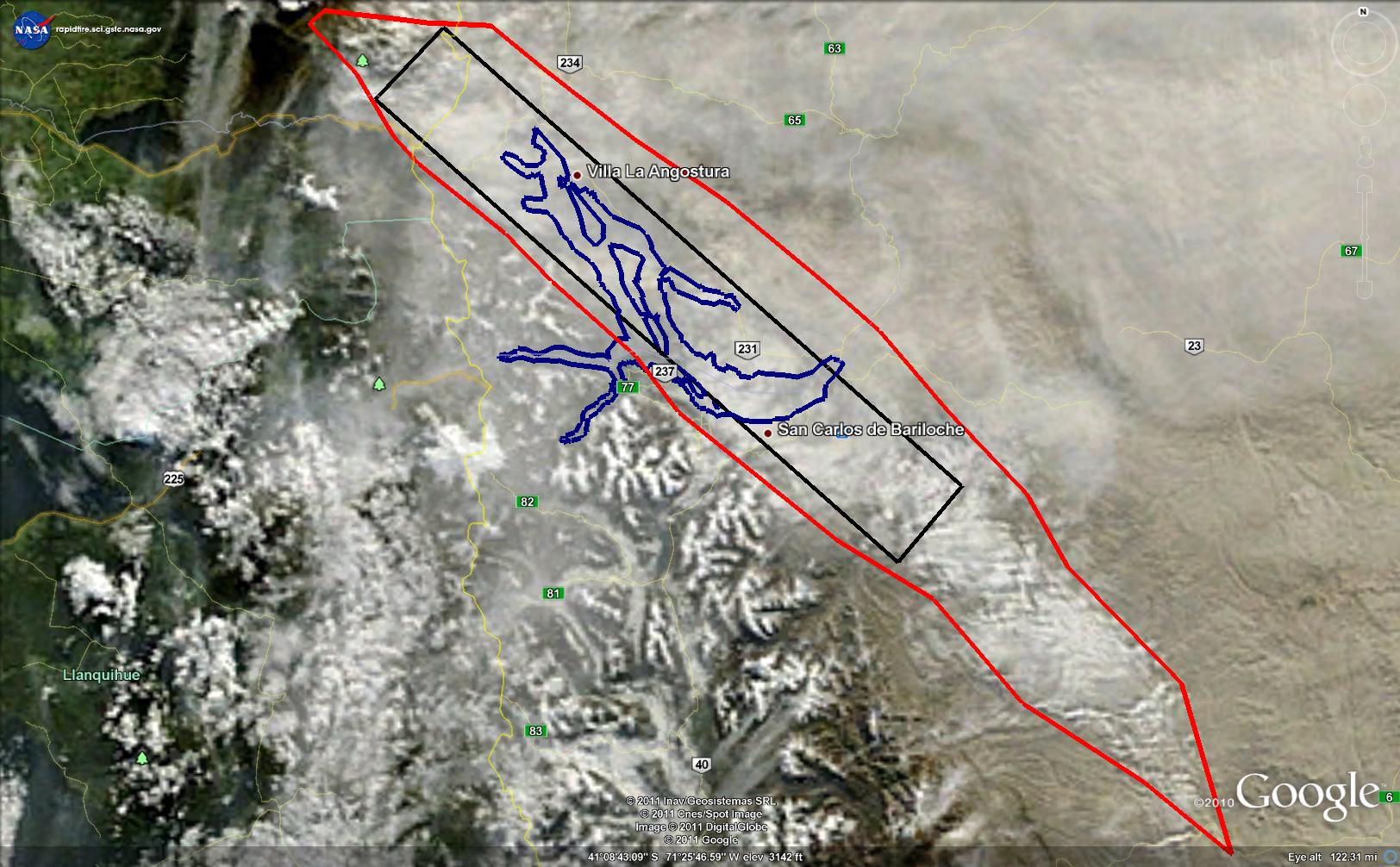}
\caption{Terra Satellite showing in red the approximated area affected by the
ashes of the first phase of the eruption. As a reference, the Nahuel Huapi lake
coastline is indicated in blue, and the black rectangle correspond to three
times the lake area.}\label{terra}
\end{figure}

In Figure\,\ref{terra} the NASA Terra satellite image for 05 Jun 2011 at 13:45
UTC is shown. The blue profile corresponds to the coastline of the lake, and in
red the area covered by sand and ashes is highlighted. As a guide to the eye, a
rectangle of area $3 A_{NH}$ is also indicated. To determine $A$, the following
data was used: the surface of the lake is $529$\,km$^2$; adding the surfaces of
the largest islands, Victoria and Huemul, totalling $32$\,km$^2$, we find
$A_{NH}=571$\,km$^2$. So the area covered by ash would be $3\times 571$\,km$^2
\simeq 1700$\,km$^2=1.7\times 10^{9}$\,m$^2$. The picture has been post
processed by using the Google Earth software\cite{gearth}.

\item Thickness of the sand layer.

In line with the above hypothesis on the covered area, we assign a value to the
average thickness of the layer of sand and ash. It is known that a layer over
$0.3$\,m covered Villa La Angostura while east of Bariloche the layer was less
than $0.1$\,m thick. On the other hand reports indicate an accumulation of more
than $0.1$\,m in the town of Ingeniero Jacobacci ($69.5^\mathrm{o}$\,W,
$41.3^\mathrm{o}$\,S) 210\,km eastward from Bariloche. This corresponds to ash
and sand that did not fell into the area $A$ but which must be included in the
computation since its elevation affects the energy balance of the volcano. To
account for the total energy cost we assume an average height of $e=0.1$\,m.
covering the chosen area $A$.

\item Duration of the first eruption.

The ash fall in Bariloche started at 16:30 local time (ART, GMT-3, 19:30 UTC)
and ended five hours later, at 21:30 ART, so the total time of the eruption
first phase is 5 hours, $t=1.8\times10^4$\,s.

\item Height reached by the cloud.

The height of interest is the difference $\Delta h=h-h_{0}$ where
$h_{0}=2000$\,m is the height of the volcano. We have $\Delta h=5000$\,m, an
average value as different components of the column reached varying heights,
the lightest getting up to $12000$\,m.

\item Density of the mixture of sand and dust.

The density of the sand and ash fell in Bariloche was determined using
household items: a measuring cup used to measure the flour or sugar and a
kitchen scale. The result was $\rho=600$\,kg\,m$^{-3}$. This value, lower than
the density of water, is explained because ash and pumice, which floats on
water, are mixed with denser components that tend to sink into water.

\end{enumerate}

\section{Results}

\subsection{Volume of sand fall}

A simple volumetric calculation shows that the volume of sand and ash fall is
\begin{equation}
V=A\times e= 1.7\times 10^{9}\,\mathrm{m}^{2}\times 0.1\,\mathrm{m}=1.7\times 10^{8}\,\mathrm{m}^{3}.
\end{equation}
Considering that a cargo truck can carry about $7$\,m$^3$, the amount of sand fall is equivalent to
\begin{equation}
\frac{V_\mathrm{sand}}{V_\mathrm{truckload}} = \frac{1.7\times 10^{8}\,\mathrm{m}^{3}}{7\,\mathrm{m}^{3}} = 2.4\times 10^{7}\mathrm{truckloads},
\end{equation}
i.e., more than twenty four million trucks of sand.

An alternative way to visualise the amount of sand consists of calculating the height it would
reach if it were placed in a square area of side $L$ by stacking a pyramid of height $H$. The volume of the pyramid $V=\frac{1}{3} L^2 h = (1/6) L^3 \tan \theta$
must match the estimated volume of sand. However, as the angle $\theta$ 
of inclination of the pyramid should not exceed the ``angle of repose'' of the
sand, approximately\cite{wiki-repose} $40^\mathrm{o}$, there is a link between the height $H$ and the base $L$. We have the relations
\begin{eqnarray}
L&=&\sqrt[3]{\frac{6V}{\tan \theta}} \\ 
H&=&\frac{L}{2}\tan \theta
\end{eqnarray}
with $\tan \alpha = 0.84$. It follows
\begin{eqnarray}
L&=&1070\,\mathrm{m} \nonumber \\ 
H&=&449\,\mathrm{m}.
\end{eqnarray}
This means that the fallen sand could be stacked in a square field $1$\,km wide
(with an area of $1$\,km$^{2}$), making a $450$\,m high pyramid. As a visual
reference we note that the area of the emergent of Huemul Island, in Bariloche
is about $90$\,ha ($\approx 0.9$\,km$^{2}$), and its height certainly less than
the $450$\,m we obtained here.

\subsection{Mass of sand and ash fall}

To determine the mass of sand fall, we multiply the calculated volume by the
measured density, to obtain
\begin{eqnarray}
m_\mathrm{sand} &=& V_\mathrm{sand} \times \rho \\
&=& 1.7\times 10^{8}\,\mathrm{m}^{3} \times 6 \times 10^{2}\,\mathrm{kg}\,\mathrm{m}^{-3} \\ 
&=& 1.03\times 10^{11}\,\mathrm{kg}\\
&=& 103\,\mathrm{Mt},
\end{eqnarray}
i.e. about 100 million of metric tonnes.

\subsection{Energy released}

The energy required to raise this mass of sand to $5000$\,m can be estimated
from the potential energy acquired when it reaches its maximum height. We have 
\begin{equation}
\Delta E_p=m g \Delta h,
\end{equation}
where $g$ is the acceleration of gravity. Thus we have
\begin{eqnarray}
E &=& mg \Delta h \nonumber \\
&=& 1.03\times 10^{11}\,\mathrm{kg} \times 9.8 \frac{\mathrm{m}}{\mathrm{s}^{2}}\times 5000\,\mathrm{m} \nonumber \\ 
&=& 5.04\times 10^{15}\,\mathrm{J} = 1.2\times 10^{3}\mathrm{kt},
\end{eqnarray}
that is more than one thousand kilotons of TNT equivalent.

For comparison, we mention that the 2011 earthquake in Japan had a magnitude of
$9.0$ in the Richter scale\cite{japan1} and released an energy\cite{japan2} of
$3.9\times 10^{22}$\,J ($\approx 10^{10}$\,kt of TNT equivalent). Another
reference point is given by the total energy produced in a country. For
example, the installed electric power in Argentina\,\cite{wiki-power} is
$P_\mathrm{Arg}=26\,000$\,MW and so, the energy produced during a day is
$E_\mathrm{Arg} = P_\mathrm{Arg} \times 86400\,\mathrm{s} = 2.25\times
10^{15}$\,J. Thus the energy released by the first phase of the eruption is
equivalent to that produced in the whole country in $5.04/2.25 \simeq
2.3$\,days.

\subsection{Output speed}

The magma rises through the conduit of the volcano mixed with fumes of
molten rock and water vapour, which results in a turbulent two phase flow.
The velocity of the magma exiting the crater can be coarsely estimated from the
energy conservation law, since the flow, at least for the denser components,
is ballistic. The ashes reaching the top of the column probably ascend the
last section by convection. For that reason we took an average altitude of $5000$\,m, a moderate value compared to the $10000$\,m or even $12000$\,m that have been reported. Neglecting air friction, the kinetic energy of fluid at the crater must be equal to the potential energy reached in the final ascent,
\[
\frac{1}{2}mv^{2}=mg\Delta h.
\]
From this it follows that
\begin{equation}
v=\sqrt{2g\Delta h}=\sqrt{2\times 9.8\frac{\mathrm{m}}{\mathrm{s}^{2}}\times 5000\,\mathrm{m}}=313\frac{\mathrm{m}}{\mathrm{s}}
\end{equation}
that is, nearly the speed of sound in air. This high speed of the outgoing
material determines its rapid cooling, which explains the amorphous
structure and the presence of micro crystals in the ejecta since production
of larger crystals requires slow crystallisation.

\subsection{Power}

Assuming that the first eruption lasted $5$\,hours, i.e. $t=1.8\times10^4$\,s, it is possible to estimate the power generated by the volcano, which turns out to be 
\begin{equation}
P=\frac{E}{t}=\frac{5.04\times 10^{15}\,\mathrm{J}}{1.8\times 10^{4}\,\mathrm{s}}=2.8\times 10^{11}\,\mathrm{W}.
\end{equation}
By comparing this result with the installed electric power in Argentina mentioned above, the ratio obtained is
\begin{equation}
\frac{P}{P_{Arg}}=\frac{2.8\times 10^{11}}{2.6\times 10^{10}}=12.
\end{equation}
The power developed by the volcano is $12$ times the installed power in Argentina. We can also compare our result with the worldwide installed electrical power\,\cite{wiki-world}, which is $P_W=15$\,TW$=1.5\times 10^{13}$\,W. A percentage comparison results in
\begin{equation}
\frac{P}{P_{W}}=\frac{2.8\times 10^{11}}{1.5\times 10^{13}}\times100=1.9\%,
\end{equation}
so the power developed by the Puyehue volcano is equivalent to about 2\% of the global electric power.

\section{Comments on the hypotheses}

In the problem of piano tuners, Fermi could compare the estimated result with
the answer obtained from a query in the telephone directory. The discrepancy
could be attributed to the fact that some tuners had no phone or were not
listed as such in the directory, or to  an error in the various estimates. But
this last point is part of the game: the idea is not to obtain an accurate
result, but to reach a reasonable approximation. As such we understand a
discrepancy not exceeding a certain limit, say half an order of magnitude, or a factor of five.

In what follows we comment on each of the scenarios we used:

\begin{enumerate} \item Area covered.

In the eruption of June 4, 2011 ash and sand covered an area larger than the
one assumed in this work. That could mean that we underestimate the energy
and power of the volcano. We tried to compensate for material falling beyond
the chosen zone, by assuming a uniform thickness of $10$\,cm in the reference
area.

\item Average thickness of the layer of sand and ash.

It is known that in the vicinity of the volcano on the Chilean side and in
Villa La Angostura the layer of sand and ash reached $40$\,cm, while further
east it did not exceeded $5$\,cm. We believe that the figure of $10$\,cm
distributed uniformly over an area equal to three times the area of the Nahuel Huapi lake (over
$1700$\,km$^2$) is appropriate as an average estimate.

\item Estimated time of activity.

The fall of sand and ashes in Bariloche ceased at 00:30 UTC (21:30 ART), that is the first eruption lasted five hours. The thunder caused by electric shock lasted a total of approximately $12$\,hours indicating that volcanic activity continued until then. This does not affect the estimation of the energy developed in the first five hours. 

\item Height reached by the cloud.

Reports on June 4 indicated a height of the column of up $12000$\,m above sea level. As the column included different density components, we believe our assumption $\Delta h=5000$\,m for the thicker components of the plume is a reasonable estimate for our purposes.

\item Density of the mixture of sand and ashes.

The density measurement was made with a sample collected within hours of the
onset of the rash. Samples taken a few days later gave higher density
values, probably because rain water dissolved or dragged away part of the components and swept them away.

\item Output speed.

We are not aware of measurements of the output speed of the magma in the
volcano being made, but specialised authors have studied the supersonic flow
of magma in volcanoes in some typical situations of the so-called explosive
pyroclastic eruptions (see for example \cite{Woods,Mitchell}), as the Puyehue-Cord\'{o}n Caulle eruption of June 4, 2011. An eruption of this type occur in the
year 79 of the Christian era at the Vesuvian volcano, inducing the tragic
disappearance of the city of Pompeii. This eruption had two distinct phases:
first a {\it{plinian}} phase, where material was ejected in a tall column, spread
in the atmosphere and fell to earth like rain; followed by a {\it{pel\'{e}an}}
phase where material flowed down the sides of the volcano as fast-moving
avalanches of gas and dust, called pyroclastic flow (pyroclasts are rock
fragments formed by a volcanic explosion or ejected from a volcanic vent).
\end{enumerate}

\section{Conclusion}

The solution of the present problem indicates that using knowledge normally
available to high school students, and introducing a series of
reasonable assumptions, it is possible to solve approximately a problem
resulting from a natural event that affects everyday life.

Moreover a comparison of the energy and power calculated with the energies
associated with human activities shows the immense magnitude of the energy
put into play in geological phenomena.

For more detailed information about the eruption we recommend reading a
detailed description of this event in references\,\cite{Bermudez1,Bermudez2}.

\section*{Acknowledgments}
We appreciate the valuable contributions and suggestions from colleagues at
UNRN and CAB-IB and the exchange of correspondence with Adriana Berm\'{u}dez and
Daniel Delpino.

As an intention to leave this article as it was sent to the arXiv
server\,\cite{Asorey2011} a few months after the eruption, we will include here
our great acknowledge to all those bloggers, web sites managers, journalists
and general public who echoes the importance of this Fermi approach to this
highly complex problem. It can be appreciated by a simple google search (such
as fermi problem volcano).

H.A. acknowledge to Prof. Luis A. N\'u\~nez for his permanent advice to publish
this work.

\end{document}